%% file: main.tex
\documentclass{article}

\usepackage{spconf,amsmath,amsfonts,graphicx}
\usepackage{amssymb,textcomp,mathtools}
\usepackage{bm,upgreek,algorithm,hyperref}
\usepackage{multirow,booktabs,hhline,array}
\usepackage{cite,url,makecell,setspace, xcolor}

\pdfoutput=1

\def\thline{\noalign{\hrule height 1.0pt}}

\renewcommand{\vec}[1]{\bm{\mathrm{#1}}}

\title{Ultra-Lightweight Speech Separation via Group Communication}
\name{Yi~Luo, Cong~Han, Nima~Mesgarani}
\address{
  Department of Electrical Engineering, Columbia University
}

\begin{document}
\ninept
\maketitle

\begin{abstract}
\input{abstract}
\end{abstract}
\noindent\textbf{Index Terms}: Speech separation, lightweight, group communication

\section{Introduction}
\input{introduction}
\label{sec:introduction}

\section{Group Communication}
\input{groupcomm}
\label{sec:GroupComm}

\section{Experiment configurations}
\input{config}
\label{sec:config}

\section{Results and discussions}
\input{result}
\label{sec:result}

\section{Conclusion}
\input{conclusion}
\label{sec:conclusion}

\section{Acknowledgments}
This work was funded by a grant from the National Institute of Health, NIDCD, DC014279; a National Science Foundation CAREER Award; and the Pew Charitable Trusts.

\bibliographystyle{IEEEtran}
\bibliography{refs}

\end{document}

%% file: abstract.tex
Model size and complexity remain the biggest challenges in the deployment of speech enhancement and separation systems on low-resource devices such as earphones and hearing aids. Although methods such as compression, distillation and quantization can be applied to large models, they often come with a cost on the model performance. In this paper, we provide a simple model design paradigm that explicitly designs ultra-lightweight models without sacrificing the performance. Motivated by the sub-band frequency-LSTM (F-LSTM) architectures, we introduce the group communication (GroupComm), where a feature vector is split into smaller groups and a small processing block is used to perform inter-group communication. Unlike standard F-LSTM models where the sub-band outputs are concatenated, an ultra-small module is applied on all the groups in parallel, which allows a significant decrease on the model size. Experiment results show that comparing with a strong baseline model which is already lightweight, GroupComm can achieve on par performance with 35.6 times fewer parameters and 2.3 times fewer operations.

%% file: introduction.tex
Despite the recent success of deeper architectures on the task of speech separation and enhancement, the model size and complexity still remain the biggest challenges in the deployment of such systems on low-resource platforms such as mobile and hearable devices. It is therefore necessary to develop novel methods for either lightweight model design or model compression while maintaining the model performance.

Various efforts have been made on designing lightweight models that have small model size and can be run in real-time. Conventional model designs in the past years include deep LSTM models with relatively large number of hidden units in each LSTM layer \cite{kolbaek2017multitalker, isik2016single, luo2017speaker}, which typically lead to large networks with tens of millions of parameters and high model complexity. Recent models have investigated the use of improvements on CNN/RNN operations \cite{yin2020phasen, hu2020dccrn}, better network architectures or organizations \cite{pandey2019new, luo2019conv, luo2020dual, tzinis2020sudo}, and fusion of convolutional and recurrent networks \cite{sun2017multiple, zhao2018convolutional, tan2018convolutional}, for creating more compact and less complex models. Neural architecture search (NAS) methods have also been utilized to automatically search for network architectures for both better performance and fewer parameters \cite{mazzawi2019improving, hu2020neural}. However, the minimum model size among all such proposals, which is around one million parameters \cite{westhausen2020dual}, is still challenging for deployments on hearable devices. On the other hand, model compression techniques have been extensively studied in the deep learning community, and methods for network pruning \cite{luo2017thinet}, distillation \cite{hinton2015distilling}, binarization and quantization \cite{hubara2017quantized, fedorov2020tinylstms}, and combination of all methods \cite{han2015deep}, have been successfully applied in various architectures. However, such operations typically introduce different levels of degradation on the model performance, and the tradeoff between the complexity and performance drop needs to be carefully considered. Many methods also require the pre-training of a large model, which may greatly introduce the overall training cost.

On the other hand, sub-band and multi-band networks have been investigated on speech enhancement and separation tasks \cite{wang2017joint, zhang2017deep, li2019multichannel}. Such methods typically process different frequency bands independently, and the same sub-network can be shared across all frequency bands. Specifically, frequency-LSTM (F-LSTM) model has been investigated for the task of automatic speech recognition (ASR) \cite{li2015lstm}. In an F-LSTM model, the full-band spectral feature is split into multiple sub-band features, and an LSTM layer is applied to learn the inter-band dependencies. The sub-band outputs are then concatenated to form the transformed full-band feature and passed to additional processing modules. However, the concatenation of the sub-band features prevents the following modules from using a small architecture, thus the overall model size in such proposals is still too large. In other models where time and frequency dependencies are jointly considered \cite{li2016exploring, sainath2016modeling, xu2018single}, either the feature concatenation is still a barrier to lightweight model design or the computational complexity can be high due to the difficulty in model parallelization across the two dimensions.

In this paper, we propose \textit{group communication (GroupComm)}, a simple extension to the F-LSTM-based methods for designing ultra-lightweight models that can be applied in any architectures. In a processing module equipped with GroupComm, the large $N$-dimensional input feature vector is split into $K$ groups where each group contains a small $M$-dimensional feature vector. A GroupComm module is then applied across all the small feature vectors for the \textit{communication} between the groups. Different from the standard F-LSTM model, we do not perform concatenation on the GroupComm output. A small processing module is then applied on the groups in parallel, and the outputs are finally concatenated to form the final output. Multiple such GroupComm-equipped processing modules can be stacked to form a deep architecture. Experiment results show that by finding the balance between the number of groups $K$ and the number of stacked processing modules, GroupComm-equipped model can achieve on par performance as the standard model in the noisy reverberant speech separation task with 35.6 times fewer parameters and 2.3 times fewer multiply-accumulate (MAC) operations \cite{whitehead2011precision}, a common metric for evaluating model complexity. This allows GroupComm to be an effective method for designing ultra-lightweight models without sacrificing the performance.

The rest of the paper is organized as follows. Section~\ref{sec:GroupComm} introduces the proposed GroupComm model design method. Section~\ref{sec:config} provides the experiment configurations. Section~\ref{sec:result} discusses the experiment results. Section~\ref{sec:conclusion} concludes the paper.

\begin{figure*}[!ht]
	\small
	\centering
	\includegraphics[width=1.6\columnwidth]{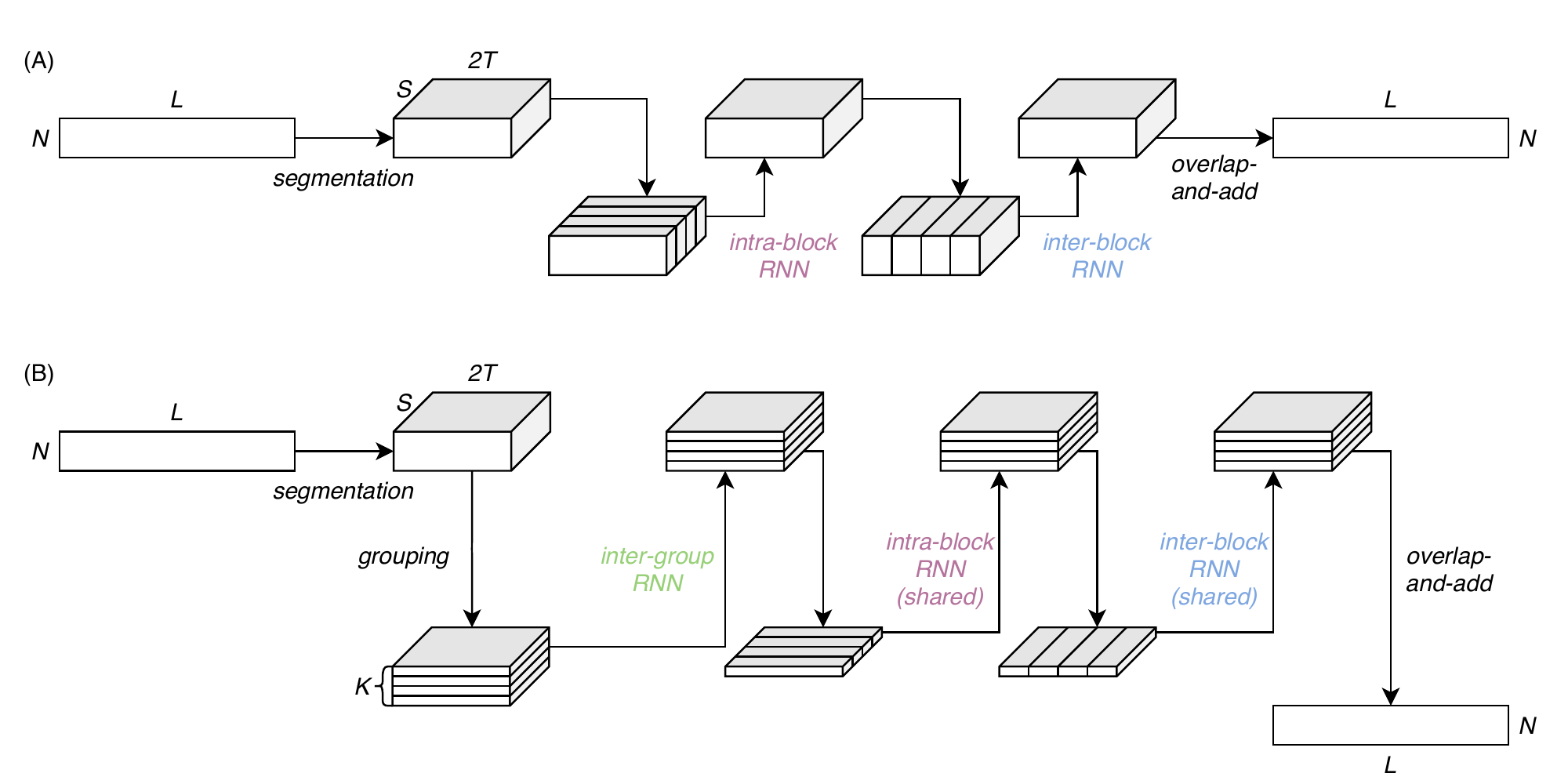}
	\caption{Flowcharts for (A) standard DPRNN block, which includes the segmentation stage, the intra-block RNN stage, the inter-block RNN stage, and the overlap-and-add stage, and (B) GroupComm-DPRNN block, which adds an additional grouping stage after the segmentation stage and applies another inter-group RNN across all the groups. The intra-block and inter-block RNNs are smaller than the ones in the standard DPRNN blocks and are shared across all groups.}
	\label{fig:flowchart}
\end{figure*}

%% file: groupcomm.tex
\subsection{Group communication}

Group communication (GroupComm) can be applied on any feature vectors. \footnote{We do not consider overlap between groups for simplicity, although any overlap ratio between the groups can be applied here.}Given a feature vector $\vec{h} \in \mathbb{R}^N$ where $N=KM$ with $K, M \in \mathbb{Z}^+$, $\vec{h}$ can be decomposed into $K$ groups $\{\vec{g}^i\}_{i=1}^K$ with $\vec{g}^i \in \mathbb{R}^M$. A network module is applied on $\{\vec{g}^i\}_{i=1}^K$ for modeling the inter-group dependencies:
\begin{align}
	\{\hat{\vec{g}}^i\}_{i=1}^K = \mathcal{F}(\{\vec{g}^i\}_{i=1}^K)
\end{align}
where $\hat{\vec{g}}^i \in \mathbb{R}^P$ is the transformed feature vector for group $i$, and $\mathcal{F}(\cdot)$ is the mapping function defined by the network module. Each $\hat{\vec{g}}^i$ is the passed to a shared module for upcoming procedures. 

\subsection{Case study: GroupComm with LSTM and dual-path RNN}

Although any network modules can be used for GroupComm, here we use LSTM and dual-path RNN (DPRNN) \cite{luo2020dual} module as a case study. Figure~\ref{fig:flowchart} provides the illustration for the standard DPRNN block and DPRNN block with GroupComm.

\subsubsection{DPRNN recap}

For a sequence of feature vectors $\vec{H} \in \mathbb{R}^{N\times L}$ where $L$ denotes the number of frames, DPRNN first segments the sequence into overlapping blocks $\{\vec{D}_i\}_{i=1}^S \in \mathbb{R}^{N\times 2T}$ of length $2T$ with hop size $T$. The first and last blocks are zero-padded accordingly in the front and at the end, respectively, so that each frame of $\vec{H}$ appears and only appears in 2 blocks. All blocks are then concatenated to form a 3-D tensor $\vec{T} = [\vec{D}_1, \ldots, \vec{D}_{S}] \in \mathbb{R}^{N\times 2T\times S}$. An intra-block bidirectional LSTM (BLSTM) is applied across the second dimension of $\vec{T}$:
\begin{align}
	\vec{U} = [\mathcal{F}_{ir}(\vec{T}[:,:,i]), \, i=1, \ldots, S]
\end{align}
where $\vec{U} \in \mathbb{R}^{H\times 2T\times S}$ is the output, and $\mathcal{F}_{ir}(\cdot)$ is the mapping function defined by the intra-block BLSTM. A linear fully-connected (FC) layer is followed to transform the feature dimension of $\vec{U}$ back to $N$, and a layer normalization (LN) layer \cite{ba2016layer} is applied to normalize the output. The normalized output is finally added to the input $\vec{T}$, leading to the output of the intra-block BLSTM $\vec{Q}$. An inter-block (B)LSTM is then applied across the third dimension of $\vec{Q}$:
\begin{align}
	\vec{V} = [\mathcal{F}_{ie}(\vec{Q}[:,i,:]), \, i=1, \ldots, 2T]
\end{align}
where $\vec{V} \in \mathbb{R}^{H\times 2T\times S}$ is the output, and $\mathcal{F}_{ie}(\cdot)$ is the mapping function defined by the inter-block (B)LSTM. Similarly, an FC layer followed by an LN layer is applied on $\vec{V}$, and a residual connection is added to $\vec{Q}$ to generate the final output of the DPRNN module. After the last DPRNN module, the 3-D tensor is reshaped to the 2-D matrix with the same shape as $\vec{H}$ via overlap-and-add on all the blocks. Figure~\ref{fig:flowchart} (A) shows the pipeline for a DPRNN module.

\subsubsection{GroupComm with BLSTM}

Although any module can be used for GroupComm, here we simply use another BLSTM layer similar to the ones in DPRNN. \footnote{We drop the subscript where there is no ambiguity.}Each frame in $\vec{T}$, denoted by $\vec{t} \in \mathbb{R}^N$, is split into $K$ vectors $\{\vec{z}^i\}_{i=1}^{2T} \in \mathbb{R}^M$ satisfying $N=KM$. A BLSTM together with its FC layer and LN layer is applied on $\{\vec{z}^i\}_{i=1}^{2T}$ and transform it to $\{\vec{w}^i\}_{i=1}^{2T} \in \mathbb{R}^M$. The original tensor $\vec{T}$ is then split into $K$ groups of smaller tensors $\{\vec{w}^i\}_{i=1}^{2T} \in \mathbb{R}^{M\times 2T\times S}$. The DPRNN module is then shared across all the $K$ groups with a $K$ times fewer hidden units than that in the DPRNN without GroupComm. Figure~\ref{fig:flowchart} (B) shows the pipeline for the GroupComm-DPRNN module.

\subsubsection{Discussion and analysis}

Using a BLSTM layer for GroupComm matches the design in standard F-LSTM model, and together with DPRNN it can be compared with the design of multi-path RNN (MPRNN) recently proposed for long sequence speech separation \cite{kinoshita2020multi}. Note that unlike the MPRNN where the segmentation only happens on the time dimension and the three RNN layers are applied on different time scales, the inter-group RNN in GroupComm-DPRNN is applied on the feature dimension and only two RNN layers are applied on the time dimension. One could easily extend GroupComm-DPRNN to GroupComm-MPRNN. Different architectures can also be applied for both GroupComm and the following processing module.

The size of DPRNN depends on the input and hidden dimensions of each of the (B)LSTM layers in the DPRNN module. Suppose that for the standard model where the input to each (B)LSTM is $N$-dimensional and the hidden units in the LSTM gates is $R$-dimensional, the GroupComm-DPRNN has (B)LSTM layers with $N/K=M$-dimensional inputs and $R/K\triangleq C$-dimensional hidden units, whose size is $K^2$ times smaller than the standard one. However, large $K$ might harm the model capacity as the number of parameters might be too small, and in Section~\ref{sec:result} we will dive into the tradeoff between model size and performance. Note that the scaling factor $K$ can be different for the input and hidden unit dimensions, e.g. the hidden unit dimension can be rescaled for $K'\neq K$, while here we simply use the same scaling factor. Using overlapped groups is left as a future work as well.

%% file: config.tex
\subsection{Dataset}

We evaluate our approach on a simulated noisy reverberant two-speaker dataset \cite{luo2020end}. 20000, 5000 and 3000 4-second long utterances are simulated for training, validation and test sets, respectively. For each utterance, two speech signals and one noise signal are randomly selected from the 100-hour Librispeech subset \cite{panayotov2015librispeech} and the 100 Nonspeech Corpus \cite{web100nonspeech}, respectively. The overlap ratio between the two speakers is uniformly sampled between 0\% and 100\%, and the two speech signals are shifted accordingly and rescaled to a random relative SNR between 0 and 5 dB. The relative SNR between the power of the sum of the two clean speech signals and the noise is randomly sampled between 10 and 20 dB. The transformed signals are then convolved with the room impulse responses simulated by the image method \cite{allen1979image} using the gpuRIR toolbox \cite{diaz2020gpurir}. The length and width of all the rooms are randomly sampled between 3 and 10 meters, and the height is randomly sampled between 2.5 and 4 meters. The reverberation time (T60) is randomly sampled between 0.1 and 0.5 seconds. After convolution, the echoic signals are summed to create the mixture for each microphone.

\subsection{Model configurations}

We use the time-domain audio separation network (TasNet) with DPRNN as the separation module for all experiments. The DPRNN-TasNet contains a trainable 1-D convolutional encoder and decoder and a DPRNN-based separation module for ``mask estimation'' on the encoder output. We use 2~ms window size (filter length) in the encoder and decoder for all experiments, and use ReLU nonlinearity as the activation function for the mask estimation layer. Note that unlike the standard DPRNN where a mask estimation layer is applied on the entire DPRNN output, in the GroupComm-DPRNN we apply different mask estimation layers for different groups. The hidden size for the mask estimation layers is rescaled according to the number of groups. This design provides a further decrease on both the model size and the number of float-point operations.

We apply different sets of hyperparameters to investigate the effect of different model configurations in GroupComm. Table~\ref{tab:param} shows the hyperparameters and their notations we will use in Section~\ref{sec:result}. The group size $M$ equals to the ratio between the number of encoder filters $N$ and the number of groups $K$. The LSTM input and output dimensions $H_i$ and $H_o$ are shared across all LSTM layers in GroupComm and DPRNN modules. The model depth is denoted by $L$.

Note that the standard DPRNN-TasNet contains a linear bottleneck layer on the encoder output for dimension reduction. Here we maintain this set-up in the baseline DPRNN-TasNet, while removing the bottleneck layer in GroupComm-equipped models and directly split the encoder output into $K$ groups. The mask estimation layer is always applied on the output of the last DPRNN or GroupComm-DPRNN module, and for GroupComm-DPRNN module the outputs from all groups are concatenated before passing to the mask estimation layer. This is to keep an identical configuration on mask estimation for both models.

\begin{table}[!ht]
	\scriptsize
	\centering
	\begin{tabular}{c|c}
		\thline
		Hyperparameter & Notation \\
		\thline
		Number of groups & $K$ \\
		Group size & $M$ \\
		Number of encoder filters & $N$ \\
		LSTM input / hidden dimensions & $H_i / H_o$ \\
		Number of DPRNN modules & $L$ \\
		\thline
	\end{tabular}
	\caption{Hyperparameters and their notations.}
	\label{tab:param}
\end{table}

\subsection{Training configurations}

All models are trained for 100 epochs with the Adam optimizer \cite{kingma2014adam} with an initial learning rate of 0.001. Signal-to-noise ratio (SNR) is used as the training objective for all models. The learning rate is decayed by 0.98 for every two epochs. Gradient clipping by a maximum gradient norm of 5 is always applied for proper convergence of DPRNN-based models. Early stopping is applied when no best validation model is found for 10 consecutive epochs. No other training tricks or regularization techniques are used. Auxiliary autoencoding training (A2T) is applied to enhance the robustness on this reverberant separation task \cite{luo2020distortion}.

\subsection{Evaluation metrics}

In addition to the SI-SDR score \cite{le2019sdr} for the evaluation of the separation performance, we report the model size and the number of MAC operations as metrics for model complexity. MACs for all models are calculated by an open-source toolbox \footnote{\url{https://github.com/Lyken17/pytorch-OpCounter}}.

%% file: result.tex
\begin{table*}[!ht]
	\scriptsize
	\centering
	\begin{tabular}{c|c|c|c|c|c|c|c|c}
		\thline
		Model & $K$ & $M$ & $N$ & $H_{i}$ / $H_{o}$ & $L$ & SI-SDR (dB) & Model size & MACs (on 4-second inputs) \\
		\thline
		DPRNN & 1 & 128 & 128 & 64 / 128 & 6 & 9.0 & 2.6M & 22.1G \\
		\thline
		\multirow{11}{*}{GroupComm-DPRNN} & 2 & 64 & 128 & 64 / 128 & 4 & 9.5 & 2.6M (1.0×) & 43.4G (0.5×) \\
		\cline{2-9}
		& 4 & 32 & 128 & 32 / 64 & 4 & 9.4 & 663.0K (3.9×) & 22.4G (1.0×) \\
		\cline{2-9}
		& 8 & 16 & 128 & 16 / 32 & 4 & 8.9 & 175.5K (14.9×) & 11.9G (1.8×) \\
		\cline{2-9}
		& \multirow{4}{*}{16} & \multirow{2}{*}{8} & \multirow{2}{*}{128} & \multirow{2}{*}{8 / 16} & 4 & 8.1 & 51.9K (50.4×) & 6.6G (3.3×) \\
		& & & & & 6 & \textbf{8.9} & \textbf{73.5K (35.6×)} & \textbf{9.6G (2.3×)} \\
		\cline{3-9}
		& & \multirow{2}{*}{16} & \multirow{2}{*}{256} & \multirow{2}{*}{16 / 32} & 2 & 8.1 & 100.7K (26.0×) & 12.4G (1.8×) \\
		& & & & & 4 & 9.7 & 183.9K (14.2×) & 23.7G (0.9×) \\
		\cline{2-9}
		& \multirow{4}{*}{32} & \multirow{2}{*}{4} & \multirow{2}{*}{128} & \multirow{2}{*}{4 / 8} & 6 & 7.6 & 26.0K (100.7×) & 5.7G (3.8×) \\
		& & & & & 10 & 8.5 & 37.6K (69.5×) & 9.1G (2.4×) \\
		\cline{3-9}
		& & \multirow{2}{*}{8} & \multirow{2}{*}{256} & \multirow{2}{*}{8 / 16} & 2 & 7.9 & 38.7K (67.6×) & 7.2G (3.1×) \\
		& & & & & 4 & 8.6  & 60.3K (43.4×) & 13.2G (1.7×) \\
		\thline
	\end{tabular}
	\caption{Performance of standard DPRNN model and GroupComm-DPRNN models with various configurations. Model size and multiply-and-accumulate operations (MACs) are also reported.}
	\label{tab:result}
\end{table*}

\subsection{Separation performance}

Table~\ref{tab:result} presents the separation performance of the baseline DPRNN-TasNet and the GroupComm-DPRNN-TasNet models with various configurations. For smaller numbers of groups at $K=2$ and $4$, a better separation performance can be achieved at the cost of an on par or higher model complexity. This shows that when computational cost is not a bottleneck, GroupComm can also be applied to improve the performance. Further increasing the number of groups without increasing either the model depth or width may lead to a worse performance, and a balance needs to be found to achieve the same level of performance as the baseline. We notice that by setting $K=16$, $N=128$ and $L=6$, the model with the same performance is only 35.6 times smaller and contains 2.3 times fewer MACs than the baseline. Such an ultra-lightweight model with only \texttildelow 70K parameters can be a strong candidate for low-resource platforms without the loss on the performance. An even smaller model with $K=32$, $N=128$ and $L=6$ only contains 26K parameters with a relative performance degradation of 15\%. This proves the effectiveness of GroupComm.

We also observe from the table that increasing the width of the model can be a good way to improve the performance, as setting $K=16$, $N=256$ and $L=4$ leads to the best separation performance across all configurations with 14.2 times smaller model size and on par model complexity. The role of model width and depth is an interesting topic left to explore.

\subsection{Drawbacks and future works}

A main drawback of the current design of GroupComm is on the memory usage during training. The memory consumption linearly increases as the model depth $L$ increases, while a a larger $L$ is important for the model capacity. A large $L$ may make the training procedure slower and prevent the use of large batch size. This drawback can be tackled by adopting better architectures for both the GroupComm and separation modules. Other works left as future works include investigating the effect of the sequential order of GroupComm and processing modules, and exploring binarization or quantization together with GroupComm.

%% file: conclusion.tex
In this paper, we proposed \textit{group communication (GroupComm)}, an extension to the frequency-LSTM (F-LSTM) model for the design of ultra-lightweight models. GroupComm split a large feature vector into groups of small vectors, and applied an inter-group module to capture the cross-group dependencies. Unlike F-LSTM that concatenated all the group-level outputs and passed it to upcoming modules, GroupComm used small processing modules on all the groups in parallel to save the model size. Multiple GroupComm-equipped modules can be stacked to increase the model capacity, and the GroupComm modules were applied iteratively throughout the entire model. Experiment results showed that applying GroupComm to the DPRNN-TasNet baseline on a simulated noisy reverberant speech separation dataset can achieve on par performance with the DPRNN-TasNet baseline with 23.5 times smaller model size and 1.7 times fewer MACs. Drawbacks and future works for GroupComm were also discussed.